\documentclass{webofc}
\usepackage[varg]{txfonts}   % Web of Conferences font
\usepackage{enumitem}
\newcommand{\panco}{\texttt{PANCO2}}
\newcommand{\eg}{\textit{e.g.}}
\newcommand{\ie}{\textit{i.e.}}

\begin{document}
\title{PANCO2: a new software to measure pressure profiles from resolved thermal SZ observations}

\author{
    \firstname{F.} \lastname{K\'eruzor\'e} \inst{\ref{LPSC}}
    \fnsep\thanks{\email{keruzore@lpsc.in2p3.fr}}
    \and
    \firstname{E.} \lastname{Artis} \inst{\ref{LPSC}}
    \and
    \firstname{J.-F.} \lastname{Mac\'ias-P\'erez} \inst{\ref{LPSC}}
    \and
    \firstname{F.} \lastname{Mayet} \inst{\ref{LPSC}}
    \and
    \firstname{M.} \lastname{Mu\~noz-Echeverr\'ia} \inst{\ref{LPSC}}
    \and
    \firstname{L.} \lastname{Perotto} \inst{\ref{LPSC}}
    \and
    \firstname{F.} \lastname{Ruppin} \inst{\ref{MIT}}
}

\institute{
    Univ. Grenoble Alpes, CNRS, Grenoble INP, LPSC-IN2P3, 53, avenue des Martyrs, 38000 Grenoble, France
    \label{LPSC}
    \and
    Kavli Institute for Astrophysics and Space Research, Massachusetts Institute of Technology, Cambridge, MA 02139, USA
    \label{MIT}
}

\abstract{%
    We have developed a new software to perform the measurement of galaxy cluster pressure profiles from high angular resolution thermal SZ observations.
    The code allows the user to take into account various features of millimeter observations, such as point spread function (PSF) convolution, pipeline filtering, correlated residual noise, and point source contamination, in a forward modeling approach.
    %One of the key advantages of the code is the possibility to use binned, non-parametric pressure profiles, enabling the detection of pressure features better than smooth functions such as the traditionally used generalized Navarro-Frenk-White profile.
    %Another major upside is the performance of the software, enabling the extraction of the pressure profile and associated confidence intervals via MCMC sampling in times as short as a few minutes.
    A major advantage of this software is its performance, enabling the extraction of the pressure profile and associated confidence intervals via MCMC sampling in times as short as a few minutes.
    We present the code and its validation on various realistic synthetic maps, of ideal spherical clusters, as well as of realistic, hydrodynamically simulated objects.
    We plan to publicly release the software in the coming months.
}
\maketitle
%
% ========================================================================== %
\section{Introduction}
\label{sec:intro}

The thermal Sunyaev-Zeldovich (tSZ) effect is a powerful way to detect and study galaxy clusters at millimeter wavelengths, thanks to their imprint on the cosmic microwave background.
The amplitude of this effect is proportional to the line of sight-integrated pressure of the intracluster medium (ICM) gas, enabling studies of the thermodynamic properties of the ICM through tSZ observations.
The importance of ICM pressure distribution in cosmology (for cluster counts and tSZ power spectra analyses) and astrophysics (\eg\ to study feedback from active galactic nuclei) therefore makes resolved SZ observations highly attractive.

The NIKA2 SZ Large Program (hereafter LPSZ, \cite{mayet_cluster_2020, perotto_nika2_2021}) is a follow-up of 50 intermediate-high redshift galaxy clusters ($0.5 < z < 0.9, \, M_{500} \in [3, 11] \times 10^{14} M_\odot$).
SZ observations are performed using the high angular resolution NIKA2 camera, observing at 150 and 260 GHz with respective angular resolutions $18"$ and $11"$ \cite{perotto_calibration_2020}.
Among its goals are the measurement of the $Y_{500} - M_{500}$ scaling relation \cite{keruzore_forecasting_2021} and of the mean pressure profile of the ICM.
These studies require the extraction of individual pressure profiles for each cluster in the sample of the program.
As a consequence, a fast and robust software able to provide an accurate pressure profile measurement must be used.

In this proceeding, we present the \panco\ (Pipeline for the Analysis of NIKA2 Cluster Observations) software, that has been developed in order to fill the needs of the NIKA2 SZ Large Program by building upon many previous developments (see \eg\ \cite{adam_first_2014, adam_pressure_2015, ruppin_non-parametric_2017, ruppin_first_2018, romero_multi-instrument_2018, keruzore_exploiting_2020}).
It uses a forward modeling regression approach, well-suited to the use of ground-based millimeter observations.
Among the key advantages of \panco\ are its speed -- with the possibility to extract a pressure profile measurement in a few minutes -- and the flexibility of the modeling, which enables the use of binned pressure profiles, accounting for correlated residual noise and data processing filtering, as well as point-source contamination.
Other codes have been publicly released in recent years, offering different and complementary options (see \eg\ \cite{castagna_joxsz_2020}, offering joint modeling of SZ and X-ray signal, but no treatment of point source contamination).

% ========================================================================== %
\section{Algorithm}
\label{sec:algo}

\begin{figure}[t]
    \begin{center}
        \includegraphics[width=.91\textwidth]{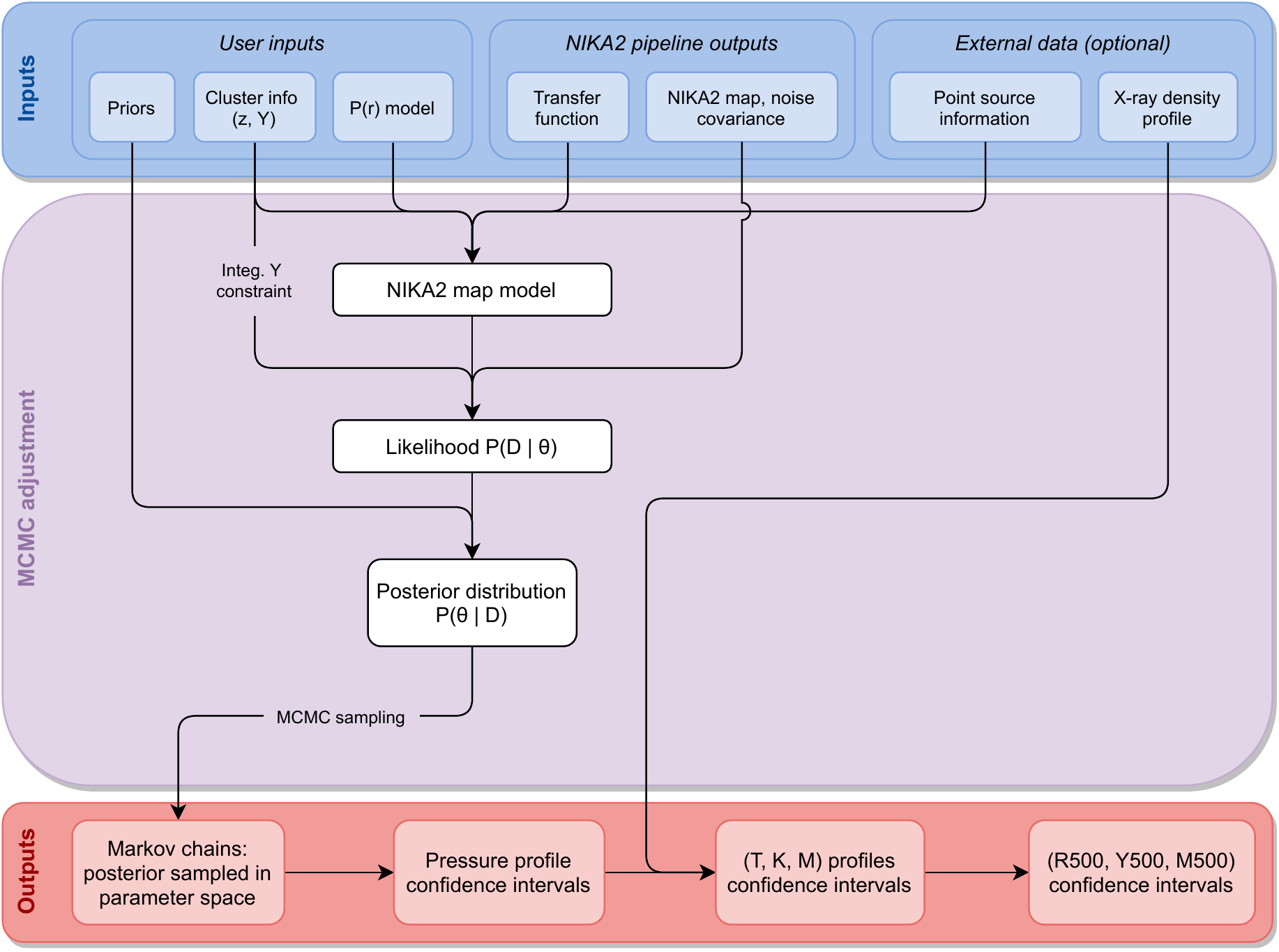}
        \caption{
            Schematic overview of the \panco\ workflow.
            The inputs (in blue) include standard SZ data reduction products, the choice of a pressure model, as well as information on point source contamination.
            The MCMC analysis yields an estimation of the ICM pressure profile, which can be combined with a density profile extracted from X-ray observations to compute the temperature $T$, entropy $K$ and hydrostatic mass $M$ profiles, as well as confidence intervals.
        }
        \label{fig:schema}
    \end{center}
    \vspace{-25pt}
\end{figure}

The main objective of \panco\ is to extract a pressure profile estimation from SZ maps, and to combine this pressure profile with a density profile extracted form X-ray observations to infer a more detailed characterization of the ICM thermodynamics.
The outline of the algorithm used by the software is presented in figure~\ref{fig:schema}.
In this section, we give a brief overview of the most relevant steps in the analysis.

% -------------------------------------------------------------------------- %
\subsection{Pressure profile modeling}

Past studies of the ICM pressure distribution have relied on very constrained pressure profile models, such as the widely-used generalized Navarro-Frenk-White (gNFW) model \cite{nagai_effects_2007}.
With the advent of deep high angular resolution observations of galaxy clusters, with instruments such as NIKA2 \cite{ruppin_first_2018}, MUSTANG2 \cite{romero_pressure_2020} or ALMA \cite{di_mascolo_joint_2019}, the number of resolution elements in SZ maps increased, enabling a finer characterization of the intracluster medium.
This has led to an increased usage of radially-binned pressure profile models, often referred to as ``\textit{non-parametric models}'' \cite{ruppin_non-parametric_2017, romero_multi-instrument_2018}.
In such parametrizations, the model parameters are the values $P_i$ of the pressure profile at pre-determined radii $R_i$, with a power-law interpolation between the bins:
\begin{equation}
    \label{}
    P(R_i < r < R_{i+1}) = P_i \left(r / R_i\right)^{-\alpha}.
\end{equation}
Using this modeling offers a larger flexibility than what can be achieved with a gNFW model, for instance by enabling the detection of localized overpressures.

\panco\ offers both gNFW and binned parametrizations of the pressure profile of the ICM, in order to let users decide for themselves what model is best-suited to their analyses, as well as to enable model comparison studies.
When choosing a binned parametrization, the radial bins $R_i$ are to be defined.
By default, they will cover the range between the instrument angular resolution and the characteristic radius $R_{500}$ of the studied galaxy cluster.

% -------------------------------------------------------------------------- %
\subsection{Forward modeling of SZ observations}
\label{sec:mapmodel}

\panco\ uses forward modeling to extract pressure profile estimates.
In this approach, the physical model -- \ie\ the ICM pressure profile -- is affected by the same operations that led to the observed data -- \ie\ a map of the SZ effect.
The forward modeling used in \panco\ can be split in 4 steps, that we detail below.

\begin{enumerate}[leftmargin=0.5cm]
\item The chosen pressure profile model is integrated along the line of sight (LoS) to create a Compton parameter $y$ map, following:
    \begin{equation}
        \label{}
        y = \frac{\sigma_\textsc{t}}{m_{\rm e}c^2} \int_{\rm LoS} P_{\rm e}(r) \, {\rm d}l,
    \end{equation}
    where $\sigma_\textsc{t}$ is the Thompson scattering cross-section, and $m_{\rm e}$ the electron mass. \\
    In the case of a gNFW profile, the line of sight integration is performed numerically.
    In the case of a binned pressure profile, we use the analytical formulation for the integration of power laws in concentric shells detailed in \cite{romero_multi-instrument_2018}\footnote{A Python implementation is publicly available on GitHub: \texttt{\url{https://github.com/CharlesERomero/MCMC_ICM_PP/blob/master/integrations/shell_pl.py}}}.

\item The Compton $y$ map is convolved with the instrumental beam to account for PSF smearing.
    If it is provided, the map is also convolved with a transfer function, to account for possible filtering in the raw data processing.

\item The resulting map is converted to surface brightness units.
    Since the conversion coefficient depends on the calibration of the maps, as well as on the beam size and atmosphere opacity at the time of observation, it is affected by uncertainty.
    It is therefore treated as a nuisance parameter in our analysis, enabling the propagation of this uncertainty to the pressure profile estimates and deriving physical properties.

\item Point source models are added to the SZ surface brightness map, with fluxes treated as model nuisance parameters.
    This approach, described in \eg\ \cite{keruzore_exploiting_2020}, allows us to account for point source contamination while also propagating uncertainties on this contamination to final physical results.
\end{enumerate}

% -------------------------------------------------------------------------- %
\subsection{MCMC analysis}

The forward modeling procedure described in the previous section produces a surface brightness map model affected by the same filtering and contaminations as the input data.
By doing so, it is possible to adjust the parameters of the pressure profile model to find the values best describing the data, through a likelihood function expressed as:
\begin{equation}
    \label{eq:like}
    -2 \log \mathcal{L}(\vartheta)
    = \big[D - \mathcal{M}(\vartheta)\big]^{\rm T}
      \Sigma^{-1}
      \big[D - \mathcal{M}(\vartheta)\big]
    + \left[\frac{Y^{\rm meas.} - Y(\vartheta)}{\Delta Y^{\rm meas.}}\right]^2,
\end{equation}
where $D$ is the observed SZ map, $\vartheta$ is a vector in the parameter space, $\mathcal{M}(\vartheta)$ the model map created from the parameters $\vartheta$, and $\Sigma$ is the noise covariance matrix.
The second term of equation (\ref{eq:like}) is a constraint on the integrated Compton parameter $Y$, which can be obtained from lower angular resolution SZ observations (\eg\ in SZ surveys).
It serves as an effective constraint on the pressure profile at scales not probed by observations.

This likelihood function is combined with priors on the model parameters to express the posterior distribution for the parameters, $P(\vartheta | D)$.
\panco\ offers a large flexibility for these priors, in order to be suited to many kinds of analyses.
The posterior distribution is sampled using a Monte Carlo Markov Chains (MCMC) algorithm, performed using the \texttt{emcee} library \cite{foreman-mackey_emcee_2019}.
Convergence of the MCMC sampling is monitored using chains autocorrelation and the $\hat{R}$ test.
The final chains are clipped to remove burn-in, and are trimmed to only keep independent points.

The Markov chains produced are used to compute the probability distribution of physical quantities of interest, starting with the pressure profile of the ICM.
If available, an electron density profile can be combined with the pressure to further characterize the thermodynamic properties of the ICM, by computing temperature, entropy, and hydrostatic mass profiles.
The latter can then be used to compute a characteristic radius $R_\Delta$ containing a mean density $\Delta$ times greater than the Universe critical density.
In turn, $R_\Delta$ is used to compute characteristic quantities integrated within the radius value, such as the integrated Compton parameter $Y_\Delta$ and the cluster hydrostatic mass $M_\Delta^{\rm HSE}$.

% ========================================================================== %
\section{Validation on simulated input}

The validation of \panco\ is performed on synthetic maps of clusters with known pressure profiles, allowing us to compare the final analysis product -- \ie\ the pressure profile reconstructed from the SZ map -- to the true underlying cluster properties.

% -------------------------------------------------------------------------- %
\subsection{Simple spherical simulations}

The first step of our validation is running \panco\ on a simulated NIKA2 map of spherical mock clusters.
We model clusters representative of the low-mass, high redshift and high-mass, low-redshift part of the NIKA2 SZ Large Program sample \cite{mayet_cluster_2020, perotto_nika2_2021}.
We generate realistic gNFW pressure profiles for both of these clusters.
We then use the forward modeling algorithm described in section~\ref{sec:mapmodel} to create realistic maps from their pressure profiles, including NIKA2 beam smearing and transfer function filtering, and realistic -- albeit uncorrelated -- residual noise.
These maps are cropped to the NIKA2 field of view, corresponding to the size of the largest angular scales that are recovered from NIKA2 cluster observations.
We run \panco\ on the resulting maps to reconstruct the clusters pressure profiles, and compare these results with the known true profile of each cluster.

\begin{figure*}[t]
    \centering
    \includegraphics[height=4.5cm]{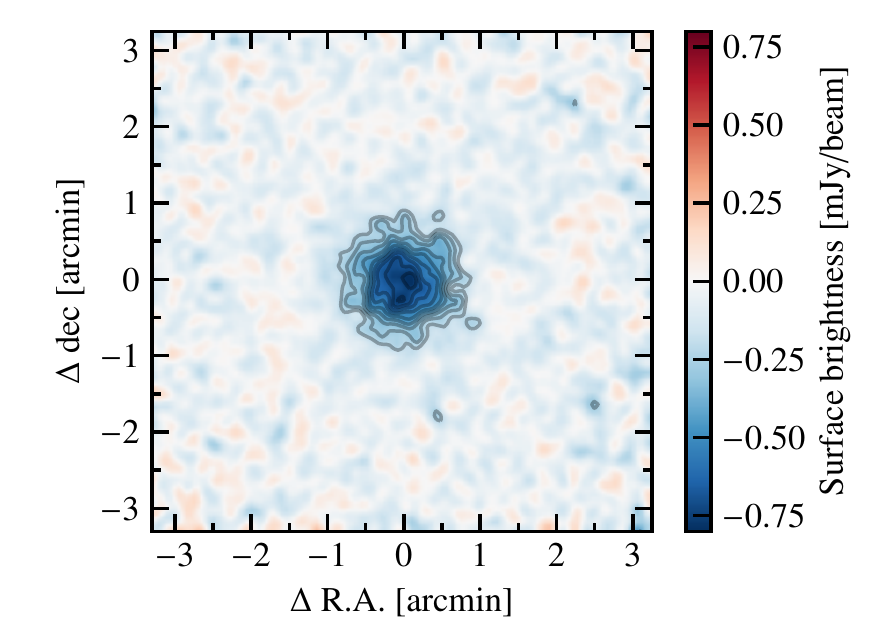} \hspace{5pt}
    \includegraphics[height=4.5cm]{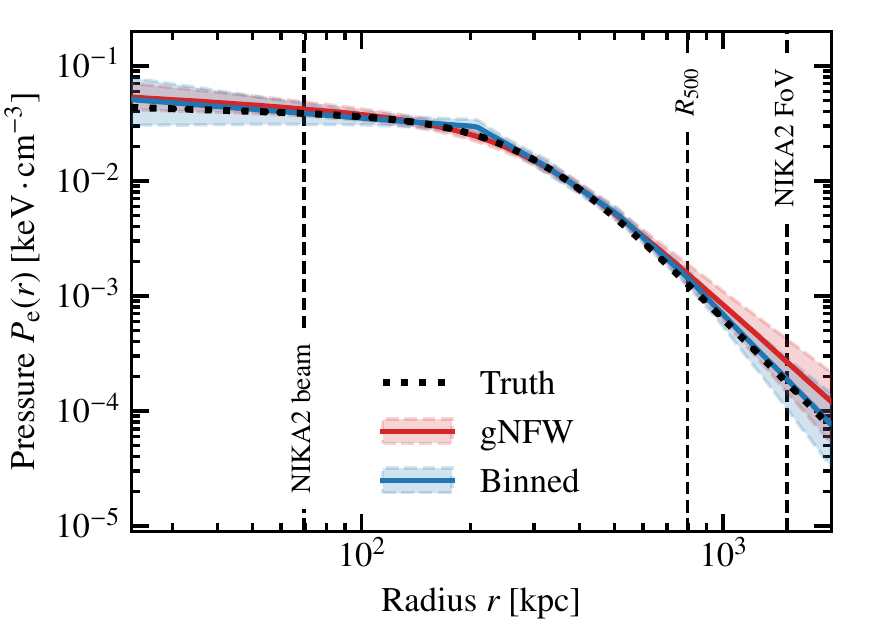}
    \caption{
        \textbf{Left:} Simulated NIKA2 2 mm surface brightness map for a low mass, high redshift galaxy cluster.
        Grey contours show signal-to-noise levels starting from $3\sigmaup$ with $1\sigmaup$ spacing.
        \textbf{Right:} Pressure profile estimates obtained by running \panco\ on the simulated map.
        Both the gNFW (red) and binned (blue) fitted pressure models agree with the input profile (black) within $1\sigmaup$.
    }
    \label{fig:validation_analytic}
    \vspace{-15pt}
\end{figure*}

The analysis is performed on both clusters, using both a gNFW and binned pressure profile model.
We present the results for both model adjustments on the low-mass, high-redshift cluster in figure \ref{fig:validation_analytic}.
The results are very similar for each of the four studies, showing excellent agreement between recovered and true pressure profiles, from the angular resolution of NIKA2 to its field of view.
Typical running times are around 10 minutes for a binned model fit and 30 minutes for a gNFW, using 30 walkers on 30 parallel threads.
The data used for this analysis will be released with \panco.

% -------------------------------------------------------------------------- %
\subsection{Hydrodynamical simulations}

The conclusions that can be drawn from the analysis of simple, spherical clusters are limited by the fact that real clusters are rarely such ideal objects.
A more thorough validation must therefore make use of galaxy clusters with more realistic physical properties.
To that end, we use simulated observations of galaxy clusters from the MUSIC\footnotemark\ hydrodynamical simulation \cite{sembolini_music_2013}.
\footnotetext{\texttt{\url{https://music.ft.uam.es}}}
Mock NIKA2 maps of these clusters were created for a sample of 32 clusters mimicking the NIKA2 SZ Large Program in \cite{ruppin_impact_2019}.
These maps include SZ signal from the line of sight projection of the ICM pressure distribution, realistic beam and transfer function filtering, as well as correlated residual noise.
They therefore constitute an ideal test-case for \panco, and can help assess its ability to recover pressure profile estimates from realistic observations.

\begin{figure*}[t]
    \centering
    \includegraphics[height=4.5cm]{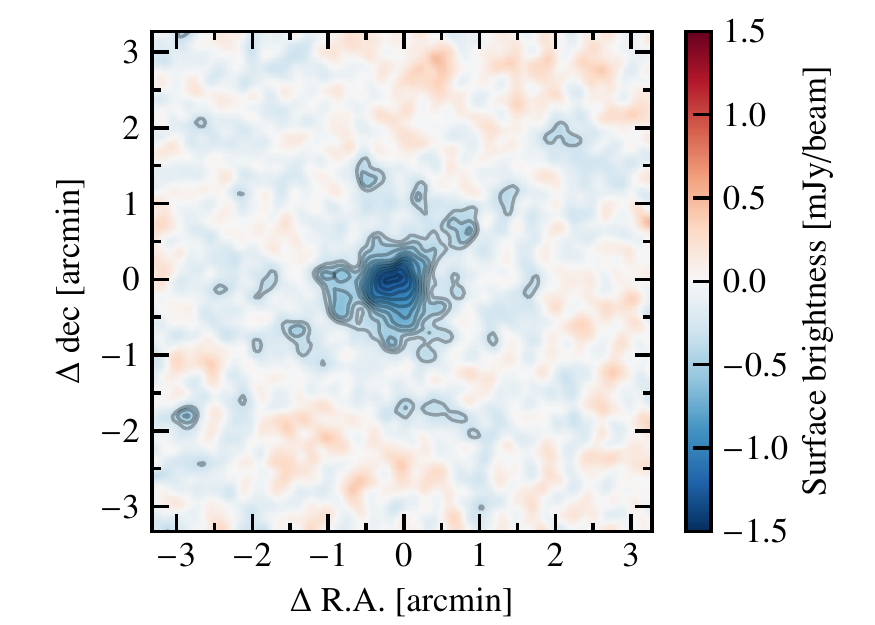} \hspace{5pt}
    \includegraphics[height=4.5cm]{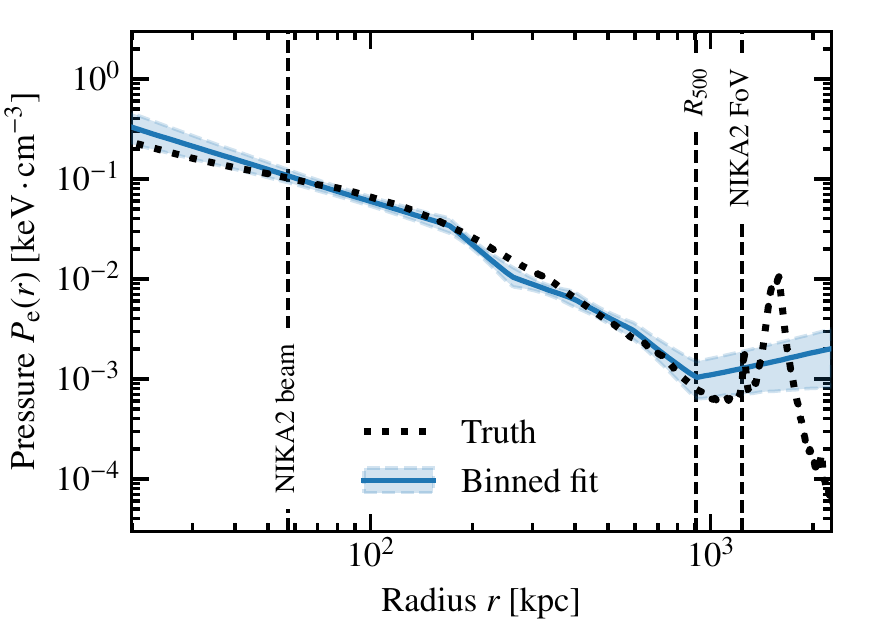}
    \caption{
        \textbf{Left:} Simulated NIKA2 2 mm map for a cluster from the MUSIC simulation at $z=0.54$.
        Grey contours show signal-to-noise levels starting from $3\sigmaup$ with $1\sigmaup$ spacing.
        \textbf{Right:} Pressure profile estimates obtained by running \panco\ on the simulated map.
        Error bars are $1\sigmaup$.
    }
    \label{fig:validation_music}
    \vspace{-15pt}
\end{figure*}

The results for one of the clusters in this sample are presented in figure \ref{fig:validation_music}.
As for the simple simulation, the recovered pressure profile is in overall agreement with the true cluster profile, with compatibility between the two profiles within $1\sigmaup$ from the NIKA2 instrumental beam to beyond its field of view.
The results are similar for the other objects of the 32 cluster sample.
This shows that \panco\ can deliver unbiased estimates of the ICM pressure profile from NIKA2 SZ observations, even for a complex pressure distribution, as expected for real galaxy clusters.

% ========================================================================== %
\section{Conclusions}

We have presented \panco, a new software to perform pressure profile measurements from thermal SZ observations.
The forward modeling MCMC approach used enables accounting for various millimeter observational artifacts to deliver unbiased pressure profile estimates and confidence intervals.
This was demonstrated using simulated cluster maps, showing good agreement between underlying and recovered pressure profiles on radial ranges covered by observations.
\panco\ is currently being adapted to be able to take as input SZ data from any instrument, so that it will be usable for all current and future millimeter surveys and pointed observations.
This software will be made public in the coming months.

% ========================================================================== %
\section*{Acknowledgements}
\small

This work is supported by the French National Research Agency in the framework of the ``Investissements d’avenir'' program (ANR-15-IDEX-02).
291294).
F.R.  acknowledges financial supports provided by NASA through SAO Award Number SV2-82023 issued by the Chandra X-Ray Observatory Center, which is operated by the Smithsonian Astrophysical Observatory for and on behalf of NASA under contract NAS8-03060.

\bibliography{./mmu_panco2.bib}

\end{document}